# Hollow Needle Puncture Mechanics for Biopsy Sampling

Yiting Wu[1], Frederic Lechenault[2], Matteo Ciccotti[2], Mattia Bacca[1*]

[1]Mechanical Engineering Department, Institute of Applied Mathematics
School of Biomedical Engineering
University of British Columbia, Vancouver BC V6T 1Z4, Canada
[2]Soft Matter Sciences and Engineering Lab, ESPCI Paris, PSL University, CNRS, Sorbonne Université, 75005 Paris, France

[*]Corresponding author: mbacca@mech.ubc.ca

**Abstract**

Biopsy sampling relies on hollow needles that puncture soft tissues by propagating and opening a cylindrical crack, yet the mechanics governing this coring process remain only partially understood. Motivated by this gap, we develop a simple, energy-based model for puncture by blunt hollow needles, grounded in brittle fracture mechanics and extended to include frictional interactions at the needle–tissue interface. The model describes puncture as the competition between the fracture energy and the elastic energy. This energetic balance is controlled by the interplay among needle geometry (radius and wall thickness), material properties (toughness and elastic modulus), and interfacial parameters (adhesion and friction). This model provides semi-analytical predictions for five key quantities: core size, frictionless force, frictional force slope, critical insertion depth, and critical insertion force. Model predictions are validated against experiments, demonstrating that friction significantly improves force estimation and alters the puncture regime. These results offer quantitative insight into the mechanics of tissue coring and force generation during biopsy, providing a predictive foundation for needle design, sampling performance, and real-time control in robotic biopsy and needle-insertion systems.

## Introduction

Puncture of soft solids is a common phenomenon in daily life, from mosquito bites [Lenau2017] to tire puncturing [Grogan1974], and plays a central role in numerous medical procedures [vanGerwen2012]. Minimally invasive interventions such as biopsy [Zhao2025], tumor resection [Huang2014], and blood sampling [Galena1992] rely on controlled needle insertion into soft tissues to access target locations. The accuracy and safety of these procedures depend critically on precise needle placement and controlled penetration forces [Abolhassani2007]. Because needle insertion governs tissue deformation, damage initiation, and sampling efficiency, a predictive understanding of puncture and coring mechanics is essential for clinical practice, biomedical device design, and robot-assisted interventions [Yang2018].

In this paper, we develop a simple mechanics-based model for hollow needle insertion and penetration, capturing the coupled roles of material properties and needle geometry in governing puncture behavior.

Most existing mechanical models of puncture in soft solids assume frictionless needle–tissue interactions, leading to a depth-independent (constant) penetration force [Shergold2004, Fregonese2021, Rattan2019, Barney2021, Barnett2016, Spagnoli2022, Fregonese2023, Montanari2023, Fregonese2022]. This assumption is also adopted in a recent model of soft coring via hollow needles [Lechenault2023], where the penetration force is derived under frictionless conditions. However, extensive experimental evidence indicates that friction plays a significant role in puncture and cutting mechanics [Fregonese2022, Goda2024, Montanari2024, Casanova2014, Asadian2014], and its implications were theoretically formalized in [Fregonese2022]. Moreover, force predictions in [Lechenault2023] rely on material properties inferred from the size of the extracted core measured within the same experiments, thereby limiting the model's predictive capability.

Here, we advance the mechanical understanding of hollow needle puncture by introducing a Griffith-type energy criterion to describe propagation of the cylindrical crack separating the core from the surrounding material. This framework enables prediction of the extracted core radius from independently measured material properties. We further incorporate frictional effects into the hollow-needle model and derive the complete force–displacement response, capturing (i) the critical force and depth at insertion and (ii) the post-insertion force evolution during penetration.

We show that friction does not significantly influence the extracted core size but plays a major role in both needle insertion and subsequent penetration. The model is validated against experiments from [Lechenault2023] as well as new experiments performed in this study. Notably, the dimensionless core size $c/R$ (with $c$ the core radius and $R$ the needle outer radius) scales linearly with the dimensionless needle radius $R/l$, where $l = \Gamma/\mu$ is a material length defined by the ratio of toughness $\Gamma$ to shear modulus $\mu$. This length characterizes the balance between fracture resistance and rigidity and correlates with the critical crack-opening scale at failure. We further demonstrate that puncture forces scale with both the needle radius $R$ and the material length $l$, supporting the intuitive result that larger coring volumes and tougher materials require greater insertion forces.

Overall, we provide a simple predictive model for hollow needle puncture, validated against experiments, that offers quantitative guidance for the design of biopsy sampling tools and for force prediction in minimally invasive procedures involving controlled material removal.

## Mechanical Model for Hollow Needle Puncture

To describe hollow needle puncture, we first analyze the forces governing the steady penetration of the needle into the specimen following insertion. We then examine the energetic requirement for penetration and compare it with the elastic energy stored during indentation. This comparison provides the critical condition for needle insertion.

### *Hollow Needle Penetration*

The penetration force of a hollow needle is due to a combination of propagation of a cylindrical crack, and the opening of the same crack to form a cavity hosting the needle inside the specimen. After insertion, the penetration force can be expressed as a linear function of the penetration depth $D$ as [Fregonese2021, Fregonese2022]

$$F_p = F_0 + F_p' D \quad (1)$$

Where $F_0$ is the frictionless penetration force, while $F_p'$ is the force-depth slope due to friction.

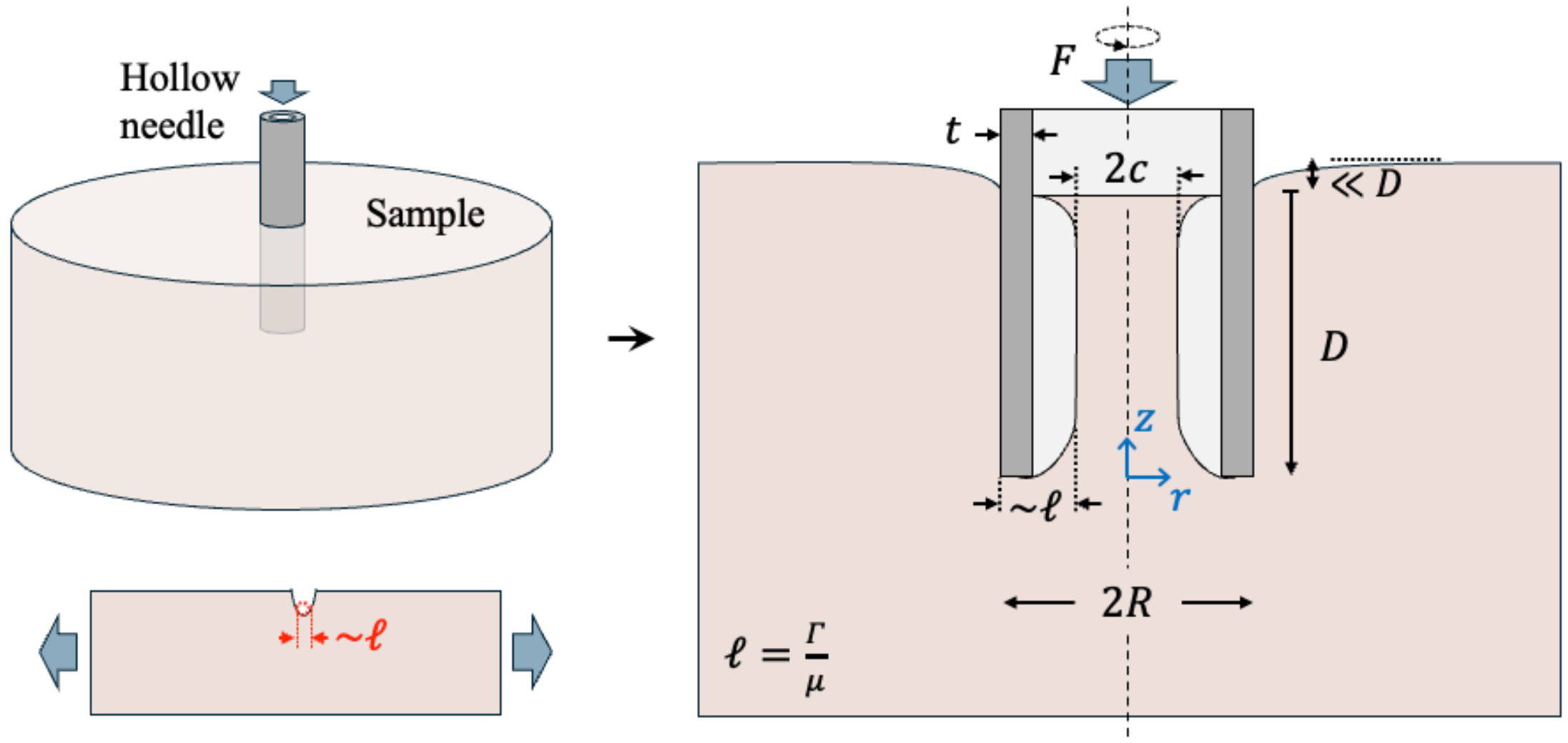

**Figure 1**: Schematics of hollow needle puncture. A hollow needle of outer radius $R$ and wall thickness $t$ pierces a soft sample, producing a cylindrical biopsy core of radius $c$. The needle is driven by an axial force $F$ and penetrates to a depth $D$. The spacing between the core and the outer surface of the needle scales with the elasto-fracture length $\ell = \Gamma/\mu$, where $\Gamma$ and $\mu$ are the toughness and shear modulus of the cored material. The bottom-left inset schematically illustrates the crack-opening radius of curvature $\ell$ in a uniaxial tensile fracture test.

*Frictionless Penetration*

The frictionless penetration force $F_0$ is defined as the rate of increment of the mechanical work $W_0$ required to penetrate the sample, i.e.,

$$F_0 = \frac{\partial W_0}{\partial D} \tag{2}$$

The total work can be written as

$$W_0 = A_c\Gamma + U_0 \tag{3}$$

where the first term represents the work required to propagate the crack, with $\Gamma$ the toughness of the material, and

$$A_c = 2\pi c D \tag{4}$$

is the crack surface in the reference configuration, with $c$ the radius of the extracted core (Fig. 1). The second term in Eq. (3) is the strain energy stored in the sample. Here, $c$ denotes the crack radius in the reference configuration, corresponding to the unloaded radius of the extracted core, whereas $R$ denotes the deformed cavity radius imposed by the needle during penetration.

Assuming a steady-state regime in which the deformation field translates with the needle, the stored energy scales linearly with penetration depth,

$$U_0 = u_0\, D \tag{5}$$

so that the core radius $c$, affecting $u_0$, can be treated as independent of $D$.

Substituting Eqs. (4) and (5) into Eq. (3), and minimizing $W_0$ with respect to $c$ at fixed depth $D$, *i.e.*, imposing $\partial W_0 / \partial c = 0$, yields

$$\Gamma = -\frac{\partial u_0}{\partial (2\pi c)} \tag{6}$$

This condition is equivalent to the Griffith criterion $G = \Gamma$ [Griffith1921], with energy release rate $G = -\partial U_0 / \partial A_c$, from Eqs. (4) and (5), taking $D$ constant. Thus, the system selects the core radius $c$ corresponding to the path of least resistance, where the elastic energy released during penetration balances the fracture energy required for crack propagation.

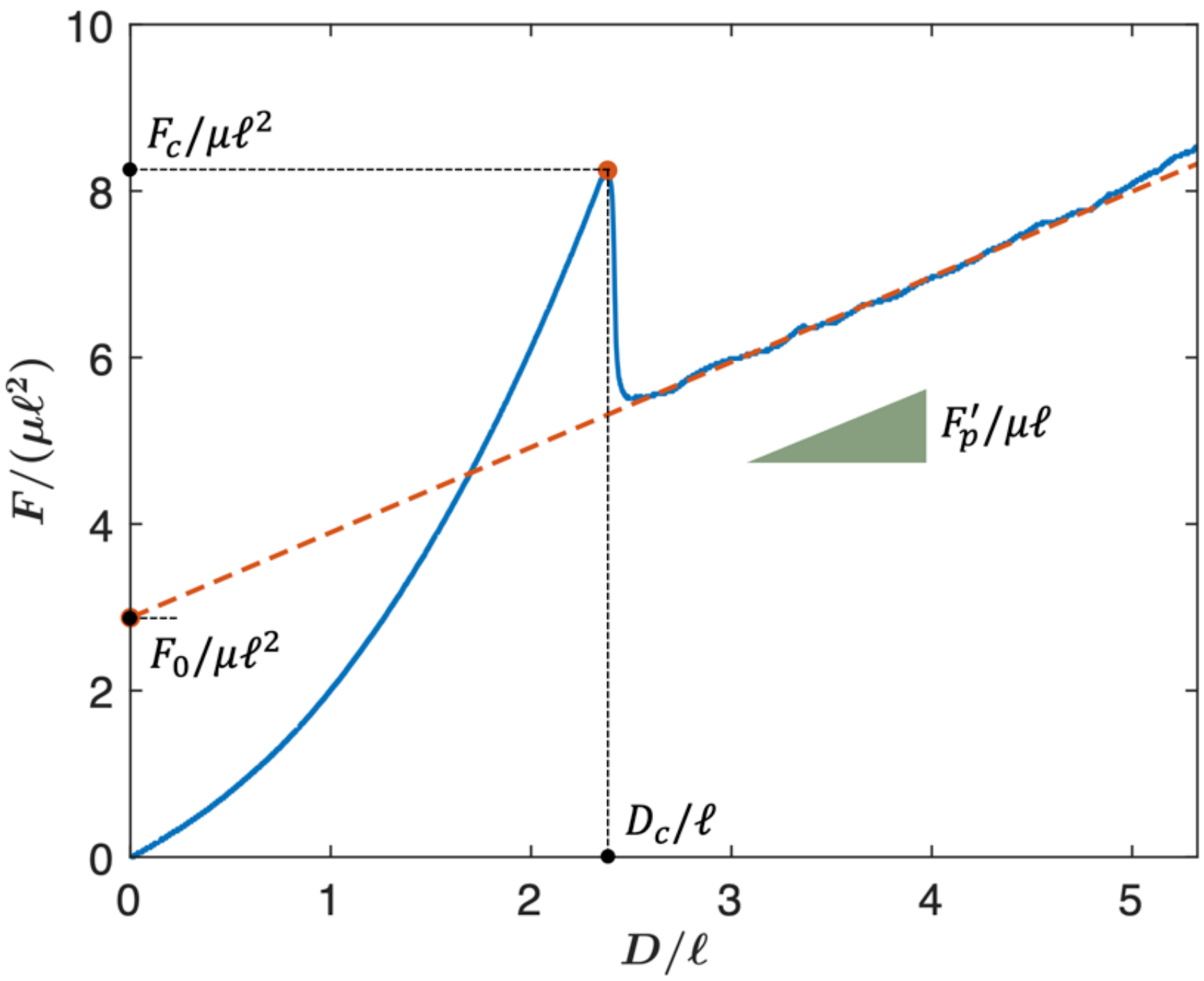

**Figure 2**: Dimensionless force-depth relation in hollow needle puncture. Like in regular puncture, the needle initially indents the material, at $D < D_c$. At insertion, $D = D_c$ and the critical maximum indentation force $F_c$ defines needle insertion [Fregonese2021]. The post-

insertion penetration force increases linearly with depth due to frictional resistance at the needle-sample interface [Fregonese2022]. At this stage, the force slope is $F_p'$, while the intercept is the frictionless puncture force $F_0$. Depth and force are both reported in the dimensionless forms $D/\ell$ and $F/\mu\ell^2$, with $\ell = \Gamma/\mu$ the elasto-fracture length, $\Gamma$ toughness, and $\mu$ the shear modulus of the material. $\ell$ represents the critical crack opening at propagation.

We describe the strain-energy density per unit reference volume of the sample using a one-term incompressible Ogden model, written in polar coordinates as

$$\varphi = \mu \, \frac{2}{\alpha^2}\left(\lambda_r^\alpha + \lambda_\theta^\alpha + \lambda_z^\alpha - 3\right) \tag{7}$$

where $\mu = E/3$ is the shear modulus, with $E$ the Young's modulus, and $\alpha$ is the strain-stiffening parameter. The parameter $\alpha$ controls the asymptotic strain-stiffening response of the material; in uniaxial tension, the engineering stress scales as a power law of the stretch at large deformation, with exponent $\alpha - 1$ [Shojaeifard2025].

The strain energy stored in the outer ring of the sample, outside the extracted core, is evaluated assuming plane-strain conditions, so that $\lambda_z = 1$, and

$$\lambda_r = \frac{\partial r}{\partial s} = \frac{1}{\lambda_\theta} = \frac{s}{r} \tag{8}$$

where $r$ and $s$ denote the radial coordinate in the current and reference configurations, respectively. Integrating Eq. (8), with the condition $r = R$ at $s = c$ at the contact with the hollow needle, gives

$$r^2 = s^2 + R^2 - c^2 \tag{9}$$

Substituting Eq. (8) into Eq. (7), the strain-energy density becomes

$$\varphi = \mu \, \frac{2}{\alpha^2}\left[\left(\frac{s}{r}\right)^\alpha + \left(\frac{r}{s}\right)^\alpha - 2\right] \tag{10}$$

The strain energy per unit penetration depth in the outer ring is then

$$u_0 = \mu\pi R^2 \, \frac{4}{\alpha^2}\int_R^\infty \left[\left(\frac{s}{r}\right)^\alpha + \left(\frac{r}{s}\right)^\alpha - 2\right]\frac{r}{R}\frac{dr}{R} \tag{11}$$

where we used $s\,ds = r\,dr$, which follows from incompressibility. After substituting Eq. (9), the integral in Eq. (11) is solved numerically following [Shergold2004]. Only for $\alpha = 2$, corresponding to the neo-Hookean case, does it admit a closed-form solution

$$u_0 = \mu\pi R^2 \left(1 - \frac{c^2}{R^2}\right)\ln\left(\frac{R}{c}\right) \tag{14a}$$

An asymptotic linear-elastic solution can also be obtained by assuming $\varepsilon_z = 0$ and $\varepsilon_r = -\varepsilon_\theta = -e$, with $\varepsilon_r = v_{,r}$ the radial strain, $v$ the radial displacement, and $\varepsilon_\theta = v/r$ the hoop strain. This gives $e = c(R - c)/r^2$, so that $\phi = 2\mu e^2$ (see at the end of *Appendix B*). Integrating the strain-energy density over the outer ring, $u_0 = \int_c^\infty \phi\, 2\pi r\, dr$, yields

$$u_o \simeq \mu\pi R^2\, 2\left(1 - \frac{c}{R}\right)^2 \tag{14b}$$

Substituting Eq. (14a) into Eq. (6) gives

$$\ell = R\, \frac{1}{2}\left[\frac{R}{c} - \frac{c}{R} + 2\frac{c}{R}\ln\left(\frac{R}{c}\right)\right] \tag{15a}$$

whose numerical solution provides the core radius $c$ for neo-Hookean materials. Substituting instead Eq. (14b) into Eq. (6) gives the linear-elastic estimate

$$c \simeq R - \frac{\ell}{2} \tag{15b}$$

where

$$\ell = \frac{\Gamma}{\mu} \tag{16}$$

is the elasto-fracture length, proportional to the critical crack-opening scale at the onset of propagation. Note that the gap between the needle outer radius and the core, $R - c$, scales with the elasto-fracture length $\ell$, as follows directly from Eq. (15b) and is illustrated in Fig. 1.

While $c$ is obtained analytically in the linear-elastic limit, it must in general be determined numerically from Eq. (15a), or equivalently from the numerical evaluation of Eq. (11) for a generic strain-stiffening parameter $\alpha$.

Only in the linear-elastic case can the frictionless penetration force $F_0$ be obtained analytically. Using Eqs. (2)–(5) and Eq. (15b), we obtain

$$F_0 \simeq \mu\ell^2\, \pi\left(2\frac{R}{\ell} - \frac{1}{2}\right) \tag{17}$$

showing that the penetration force scales linearly with the needle radius $R$.

The linear-elastic estimate of the core radius from Eq. (15b) is positive only for $R > l/2$, while the corresponding estimate of the frictionless force from Eq. (17) is positive for $R > l/4$. This inconsistency indicates that the linear-elastic approximation loses physical validity for sufficiently small needle radii, and should therefore be regarded as an asymptotic approximation applicable in the limit of large $R$.

*Frictional Resistance to Penetration*

The penetration resistance generated by interfacial friction is given by

$$F_p' = 2\pi R\tau_f \tag{18}$$

where the interfacial shear strength is

$$\tau_f = \tau_0 + fp \tag{19}$$

Here, $\tau_0$ represents an adhesive (pressure-independent) contribution to friction, $f$ is the Coulomb friction coefficient, and $p$ is the contact pressure at the needle–sample interface. The latter is obtained as $p = -(\sigma_r)_{r=R}$, where $\sigma_r$ is the radial Cauchy stress. Using radial equilibrium,

$$\frac{\partial \sigma_r}{\partial r} = \frac{\sigma_\theta - \sigma_r}{r} = \frac{1}{r}\left[\lambda_\theta \frac{\partial \varphi}{\partial \lambda_\theta} - \lambda_r \frac{\partial \varphi}{\partial \lambda_r}\right] \tag{20}$$

integration from $r = R$ to $r \to \infty$ yields $p = \int_R^\infty (\partial\sigma_r/\partial r)dr$. Substituting Eqs. (8) and (10), this gives

$$p = \mu\, \frac{2}{\alpha}\int_R^\infty \left[\left(\frac{r}{s}\right)^\alpha - \left(\frac{s}{r}\right)^\alpha\right]\frac{dr}{r} \tag{21}$$

This integral is evaluated numerically (using Eq. (9)). For $\alpha = 2$ (neo-Hookean), a closed-form solution is obtained

$$p = \mu\, \frac{1}{2}\left[1 - \frac{c^2}{R^2} + \ln\left(\frac{R^2}{c^2}\right)\right] \tag{22}$$

An equivalent expression for $p$ can be derived by noting that the strain energy $u_0$ corresponds to the work required to expand the cavity from radius $c$ to $R$, *i.e.*, $u_0 = \int_c^R p2\pi R dR$, so that $p = (\partial u_0/\partial R)/2\pi R$, which recovers Eq. (22) when using Eq. (14a). In the linear-elastic limit, using Eq. (14b), this procedure gives

$$p \simeq \mu\, 2\left(1 - \frac{c}{R}\right) \tag{23}$$

The frictional contribution to the penetration force is then obtained from Eq. (18) by substituting Eq. (19). In the linear-elastic approximation, using Eq. (15b) for the core radius, we obtain $p \simeq \mu l/R$, and hence

$$F_p' \simeq \mu\ell\, 2\pi\left(f + \frac{\tau_0}{\mu}\frac{R}{\ell}\right) \tag{24}$$

This expression shows that the frictional contribution to penetration scales linearly with the needle radius $R$, with the pressure-independent term $\tau_0$ controlling this dependence.

***Hollow Needle Insertion***

The insertion of a needle is characterized by an elastic instability [Fregonese2021], occurring when the strain energy stored in the sample during deep indentation becomes sufficient to activate the penetration mechanism. The indentation force is written as

$$F_i = 2\beta_1 \mu R D + 3\beta_2 \mu D^2 \tag{25}$$

where $D$ is the indentation depth, and $\beta_1$ and $\beta_2$ are coefficients that depend on the indenter tip geometry and the hyperelastic material response. Linear elasticity predicts a linear force–depth relation at small depth, giving the first term in Eq. (25), while, as observed in [Fakhouri2015, Shojaeifard2025], at large depth the response becomes parabolic, giving the second term.

The strain energy stored during indentation is given by

$$W_i = \int_0^D F_i dD \tag{26}$$

while the energy required to sustain penetration is

$$W_p = \int_0^D F_p dD \tag{27}$$

Substituting Eq. (25) into Eq. (26) gives

$$W_i = \beta_1 \mu R D^2 + \beta_2 \mu D^3 \tag{28a}$$

while substituting Eq. (1) into Eq. (27) gives

$$W_p = F_0 D + \frac{1}{2} F_p' D^2 \tag{28b}$$

The onset of needle insertion occurs when the stored elastic energy matches the energetic requirement for penetration, *i.e.*

$$W_i(D_c) = W_p(D_c) \tag{29}$$

Solving this condition yields the critical indentation depth

$$D_c = \ell \sqrt{\frac{F_0}{\beta_2 \mu \ell}} \left( \sqrt{\zeta^2 + 1} - \zeta \right) \tag{30a}$$

with

$$\zeta = \frac{\beta_1 - \frac{F_p'}{2\mu R}}{2\sqrt{\frac{\beta_2 F_0}{\mu R^2}}} \tag{30b}$$

Finally, the critical force required for needle insertion is obtained as the indentation force evaluated at the critical depth, $F_c = F_i(D_c)$, by substituting Eq. (30a) into Eq. (25). In the next section, we compare these theoretical predictions with experiments.

## Puncture and Friction Tests

To validate the model presented in the previous section, we used experimental data from [Lechenault2023] for larger hollow needles and performed additional experiments on smaller needles in this study, following the same experimental protocol. Elastomer samples were prepared using Elite Double 32 (ED32, Zhermack). The polymer and crosslinker were mixed at a 1:1 ratio by weight, degassed for 1.5–2 minutes in a vacuum chamber, and left to cure at room temperature overnight to ensure complete crosslinking.

All experiments were performed using an Instron universal testing machine (6800 series). Puncture tests were conducted with a 2 kN load cell, while friction tests used a 100 N load cell. The needles were attached to the load cell through a custom mandrel and driven at a constant crosshead speed of 10 mm/min, chosen to remain within the quasi-static regime assumed in the model.

For the puncture tests, two hollow needle sizes were used: (i) outer diameter (OD) 1.83 mm with inner diameter (ID) 1.54 mm, and (ii) OD 1.65 mm with ID 1.36 mm. Samples were placed on a wooden board to prevent needle damage upon full penetration. The needles were inserted until contact with the base plate. The resulting cores were extracted and imaged in segments using an optical microscope. Each experimental condition was repeated five times to ensure reproducibility. Core diameters were measured from the microscope images using Fiji (ImageJ). Measurements were taken along the stalk region of the core, excluding the transient clarinet-shaped ends, and averaged over multiple segments.

To measure the pressure-independent friction $\tau_0$ and the Coulomb friction coefficient $f$ appearing in Eq. (19), we performed friction tests following the approach of [Montanari2024], which consists of inserting a needle into pre-pierced rubber. While their study relied on casting samples around a cylindrical placeholder to create holes of known diameter, we generated the hole by puncturing the bulk with a hollow needle and then reinserting the same needle, as well as needles of larger diameter, to measure friction. This procedure preserves the surface roughness created during puncturing, providing a more realistic measurement of friction. A custom stage was built to allow the needle to travel beyond the sample height, ensuring a constant contact area and a stable shear-force plateau. Each needle diameter imposed a different normal pressure, and each condition was repeated five times. From the force–displacement curves, the plateau force was extracted to compute the shear stress. The radial pressure was then calculated from the measured core size using Eq. (21), and the shear–normal relationship in Eq. (19) was fitted to obtain the values of $\tau_0$ and $f$ reported in Table 1.

## Results & Discussion

In this section we compare our model predictions with experimental observations, focusing on the following puncture characteristics: core size $c$, insertion depth $D_c$, insertion force $F_c$, post-insertion force intercept $F_0$ (frictionless penetration force), and post-insertion slope $F_p'$. The material tested is Elite Double® 32 (ED32) by Zhermack. Experimental measurements of the puncture characteristics were obtained from the data reported in [Lechenault2023] (Exp. 1) and from the additional experiments performed in this study (Exp. 2).

The material parameters used in the model predictions are summarized in Table 1. The shear modulus and strain-hardening parameter were obtained from uniaxial tests reported in [Lechenault2023], using the identification method proposed in [Shojaeifard2025]. The fracture toughness was taken from [Montanari2023], while the frictional properties were measured independently using the method described in [Montanari2024].

**Table 1**: Material parameters for ED32 (Elite Double®, Zhermack). The parameters $\mu$, $\alpha$, $\beta_1$ and $\beta_2$ are extracted from uniaxial tensile tests reported in [Lechenault2023]; the fracture toughness $\Gamma$ is taken from [Montanari2023]; the elasto-fracture length $\ell$ is given by Eq. (16); the interfacial shear strength $\tau_0$ and friction coefficient $f$ are measured in the present study.

| $\mu$ [MPa] | $\alpha$ | $\Gamma$ [J/m$^2$] | $\ell = \Gamma/\mu$ [mm] | $\tau_0$ [MPa] | $\tau_0/\mu$ | $f$ | $\beta_1$ | $\beta_2$ |
|---|---|---|---|---|---|---|---|---|
| 0.38 | 2.7 | 1162 | 3.06 | 0.055 | 0.14 | 0.27 | 2.11 | 0.37 |

Fig. 3 shows the core-size prediction from the numerical solution of the hyperelastic model, using the parameters listed in Table 1, together with the linear-elastic prediction from Eq. (15b). The figure shows the relation between the dimensionless core size $c/\ell$ and the dimensionless needle radius $R/\ell$, where $\ell = \Gamma/\mu$ is the elasto-fracture length defined in Eq. (16), which scales with the critical crack opening at propagation. Both axes are logarithmic. As indicated by the slope triangles, the model predicts an approximately parabolic scaling, $c \sim R^2/\ell = R^2\mu/\Gamma$, for small needle radii, and an approximately linear scaling, $c \sim R$, for larger radii, where the hyperelastic and linear-elastic predictions converge. These trends indicate that both the shear modulus $\mu$ and needle radius $R$ increase the extracted core size, whereas the toughness $\Gamma$ reduces it. This follows from the energetic competition between propagating a cylindrical crack of radius $c$ and opening the resulting cavity from radius $c$ to the needle radius $R$. Therefore, softer and tougher materials are expected to produce a comparatively smaller core.

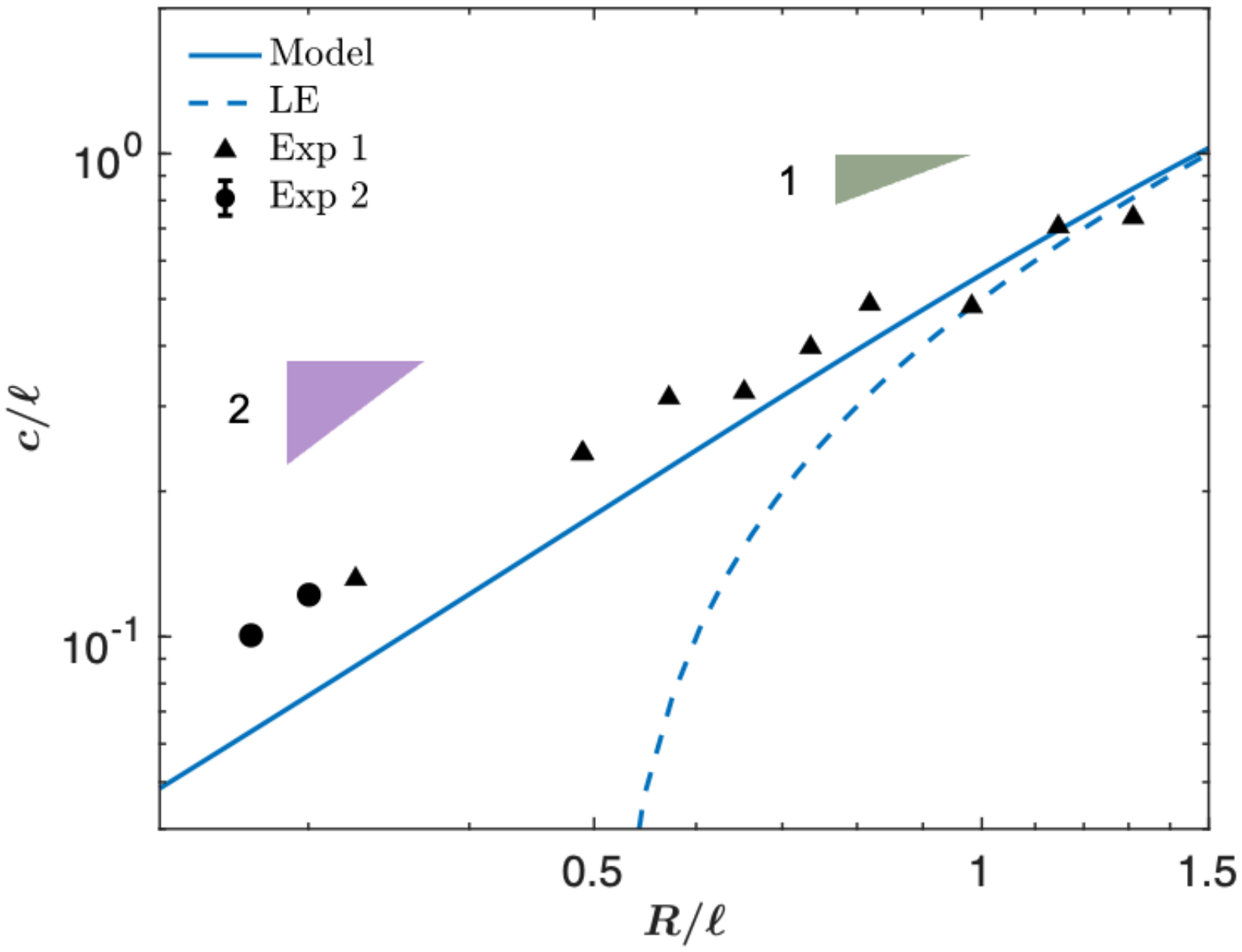

**Figure 3**: Dimensionless core size $c/\ell$ as a function of dimensionless hollow-needle outer radius $R/\ell$, where $\ell = \Gamma/\mu$ is the elasto-fracture length. Both axes are logarithmic. Predictions from the nonlinear elastic model are shown by the solid line, while the linear-elastic approximation from Eq. (15b) is shown by the dashed line. Experimental data are also shown: Exp. 1 from [Lechenault2023] and Exp. 2 from the present study.

We note that the present analysis neglects the influence of the wall thickness $t$ of the hollow needle. The mechanism described by our model applies only if the wall thickness satisfies $t < t^*$, with

$$t^* = R - c \tag{31}$$

representing the maximum wall thickness beyond which the extracted core would contact the inner wall of the needle. A thicker wall would instead activate a mechanism closer to puncture by a solid flat punch [Shergold2004], effectively suppressing the hollow-needle coring process. Eq. (31) shows that $t^*$ decreases linearly with the core radius $c$. Therefore, because increasing $R$, $\mu$, or $1/\Gamma$ tends to increase $c$, these parameters also reduce the allowable wall thickness $t^*$, although $R$ also enters Eq. (31) directly and increases $t^*$ geometrically. From this, we can observe that softer and tougher materials, which produce smaller cores, allow comparatively larger wall thicknesses before this transition occurs. From the linear elastic approximation in Eq. (15b), and (31), we can deduce that $t^* \simeq \ell/2$, also indicated in Fig. 1. Given $\ell$ is a measure of the critical crack opening displacement, this length provides an estimate of the gap opening between the core and the outer wall of the needle. For the two hollow needles used in the present experiments, the wall thickness is $t = 0.145\,mm$ (Appendix A). Using the measured core sizes reported in Fig. 3, the corresponding clearance $R - c$ is estimated to be approximately $0.3 - 0.6\,mm$, and is therefore larger than the needle wall thickness. Thus, the experimental conditions adopted in this study satisfy the geometric requirement $t < R - c$ assumed by the proposed coring model. The present experiments used flat-bottom hollow needles, corresponding to a blunt bottom-edge geometry. Tip sharpness is therefore not treated as an independent model parameter, although sharper bottom edges are expected to facilitate crack initiation and reduce the indentation depth required for insertion, whereas blunter edges increase the stored elastic energy required before coring begins.

Fig. 4 compares the model prediction with experimental estimates of the post-insertion penetration-force intercept $F_0$, obtained by extrapolating the force-depth response back to $D = 0$ (Fig. 2). In the present theory, this intercept is interpreted as the frictionless penetration force defined in Eq. (2). Since $F_0$ is not measured directly, but inferred from the post-insertion linear regime, the experimental estimates are sensitive to the choice of fitting interval, the identification of the post-insertion slope, and variability in core formation and needle-sample contact conditions. This uncertainty likely contributes to the scatter in Fig. 4, including the lower point visible in the experimental data. The parameters used in the calculation are reported in Table 1, and the core size predicted in Fig. 3 is used in the evaluation. Fig. 4 shows the relation between the dimensionless force $F_0/(\mu\,\ell^2)$ and the dimensionless needle radius $R/\ell$ on logarithmic axes. For larger radii, the trend is approximately linear, $F_0 \sim \mu\ell R = \Gamma R$, as indicated by the green slope triangle. In this regime, the hyperelastic prediction (solid line) and the linear-elastic approximation (dashed line) coincide, and the model captures the experimental trend reasonably well. At smaller radii, however, the model underestimates the experimental intercept. This discrepancy may partly arise from rate-dependent and size-dependent effects that are not included in the present quasi-static hyperelastic formulation. For a fixed insertion speed, the characteristic strain rate near the needle scales as $v/R$, with $v$ the needle velocity, so smaller needles may probe a stiffer and more dissipative material response than that described by the model. In addition, finite edge radii and fracture-process-zone sizes that are non-negligible relative to $R$ may further increase the effective penetration work. This interpretation is consistent with classical observations that strain concentration near an incision is controlled by edge sharpness [Thomas1955], and with recent work showing that blade geometry can affect deformation localization and fracture-process-zone development during cutting of soft solids [Zhan2024]. These effects would raise the measured intercept relative to the ideal frictionless prediction, particularly at smaller needle radii.

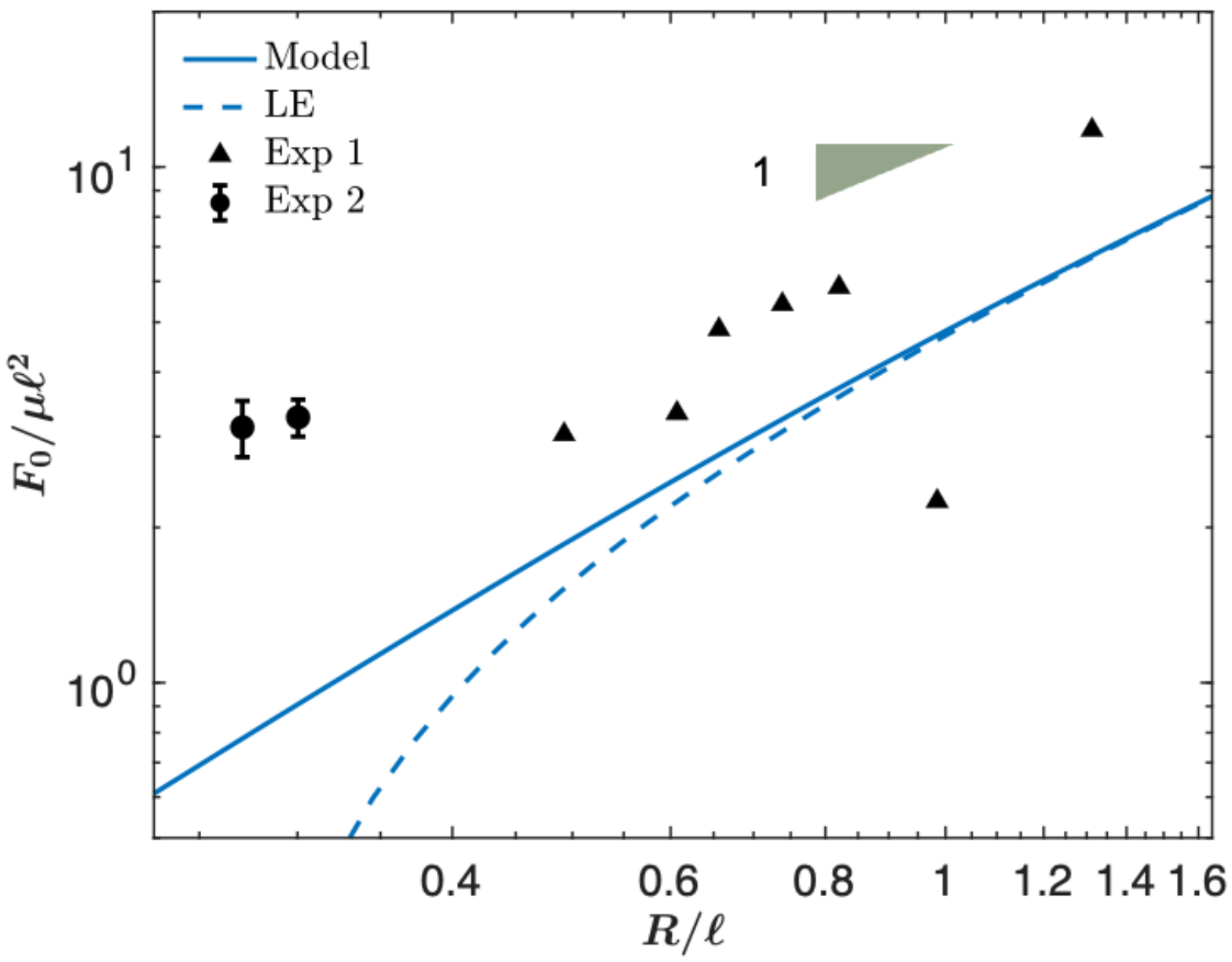

**Figure 4**: Dimensionless frictionless penetration force $F_0/(\mu\ell^2)$ as a function of the dimensionless needle radius $R/\ell$ on logarithmic axes. Predictions from the nonlinear elastic model are shown by the solid line, while the linear-elastic approximation is shown by the dashed line. The predictions are compared with experimental estimates of $F_0$ obtained from the intercept at $D = 0$ of the post-insertion penetration force response (Fig. 2). Exp. 1: [Lechenault2023]; Exp. 2: present study.

Fig. 5 shows the dimensionless post-insertion penetration-force slope $F_p'/(\mu\ell)$, introduced in Fig. 2, as a function of the dimensionless needle radius $R/\ell$. The model predictions are obtained from Eqs. (18), (19), (21), and (23), using the independently measured pressure-independent friction $\tau_0$ and Coulomb friction coefficient $f$ between the needle and the material, as described in the *Puncture and Friction Tests* section. The nonlinear elastic prediction is shown by the solid line, while the linear-elastic approximation is shown by the dashed line.

The model predicts two limiting trends. At smaller radii, the force slope follows an approximately square-root scaling, $F_p' \sim \mu\sqrt{\ell R} = \sqrt{\mu\Gamma R}$, as indicated by the orange slope triangle. In this regime, the force slope is controlled by the coupling between material stiffness, toughness, and needle radius. At larger radii, the nonlinear and linear-elastic predictions converge and approach an approximately linear scaling, $F_p' \sim \mu R$, as indicated by the green slope triangle. In this regime, the force slope becomes primarily controlled by the contact area scale and material stiffness.

Overall, the model captures the experimental trend for most needle radii, but deviations remain, especially at the largest radii. Since $F_p'$ is directly controlled by the interfacial properties $\tau_0$ and $f$, these discrepancies may reflect uncertainty in the independently measured friction parameters, or differences between the contact conditions in the reinsertion tests and those occurring during coring.

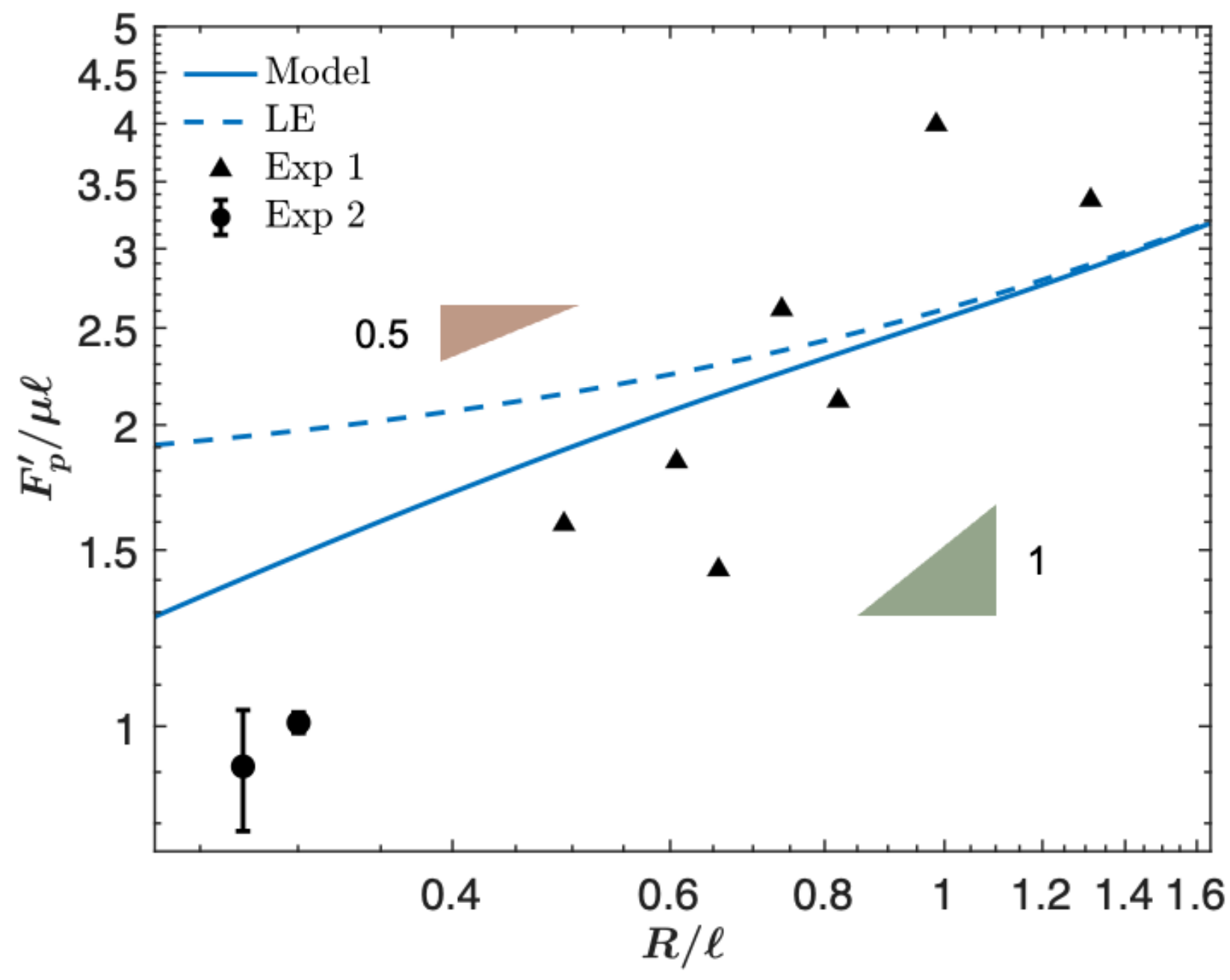

**Figure 5**: Dimensionless post-insertion penetration-force slope $F_p'/(\mu\ell)$ as a function of dimensionless needle radius $R/\ell$. Predictions from the nonlinear elastic model are shown by the solid line, while the linear-elastic approximation is shown by the dashed line. The predictions are compared with experimental data: Exp. 1 from [Lechenault2023] and Exp. 2 from the present study.

Fig. 6 plots the dimensionless critical insertion depth $D_c/\ell$ as a function of the dimensionless needle radius $R/\ell$ on logarithmic axes. Experimental measurements are compared with the nonlinear elastic model prediction (solid line) and with the corresponding linear-elastic approximation used to estimate $F_0$ (dashed line). As indicated by the orange slope triangle, the model predicts an approximately square-root scaling, $D_c \sim \sqrt{\ell R} = \sqrt{\Gamma R/\mu}$, over the range shown. Thus, the critical depth increases with toughness $\Gamma$ and needle radius $R$, but decreases with stiffness $\mu$. A similar square-root dependence of the critical insertion depth on needle radius was previously observed for solid-needle insertion [Fakhouri2015].

The model captures the overall experimental trend, although quantitative discrepancies remain. This is expected because $D_c$ is obtained from an energy balance involving both the frictionless penetration force $F_0$ and the post-insertion force slope $F_p'$; therefore, uncertainties in either quantity, including the identification of $F_0$ and the calibration of the interfacial parameters controlling $F_p'$, propagate into the prediction of $D_c$. Similar discrepancies in the prediction of critical insertion depth have also been reported for solid needles, including [Fregonese2021], which motivated the energy-based criterion used here.

An additional source of uncertainty is the assumption of near-complete surface recovery after insertion, schematized in Fig. 1 by a residual indentation scale much smaller than $D$. This approximation allows the indentation depth before insertion to be identified with the penetration depth after insertion. However, residual indentation has been observed experimentally, both here and in related measurements [Shrestha2024], and may not be negligible at the onset of insertion. This would modify both the stored indentation energy available to drive coring and the effective penetration depth, contributing to discrepancies in the prediction of $D_c$.

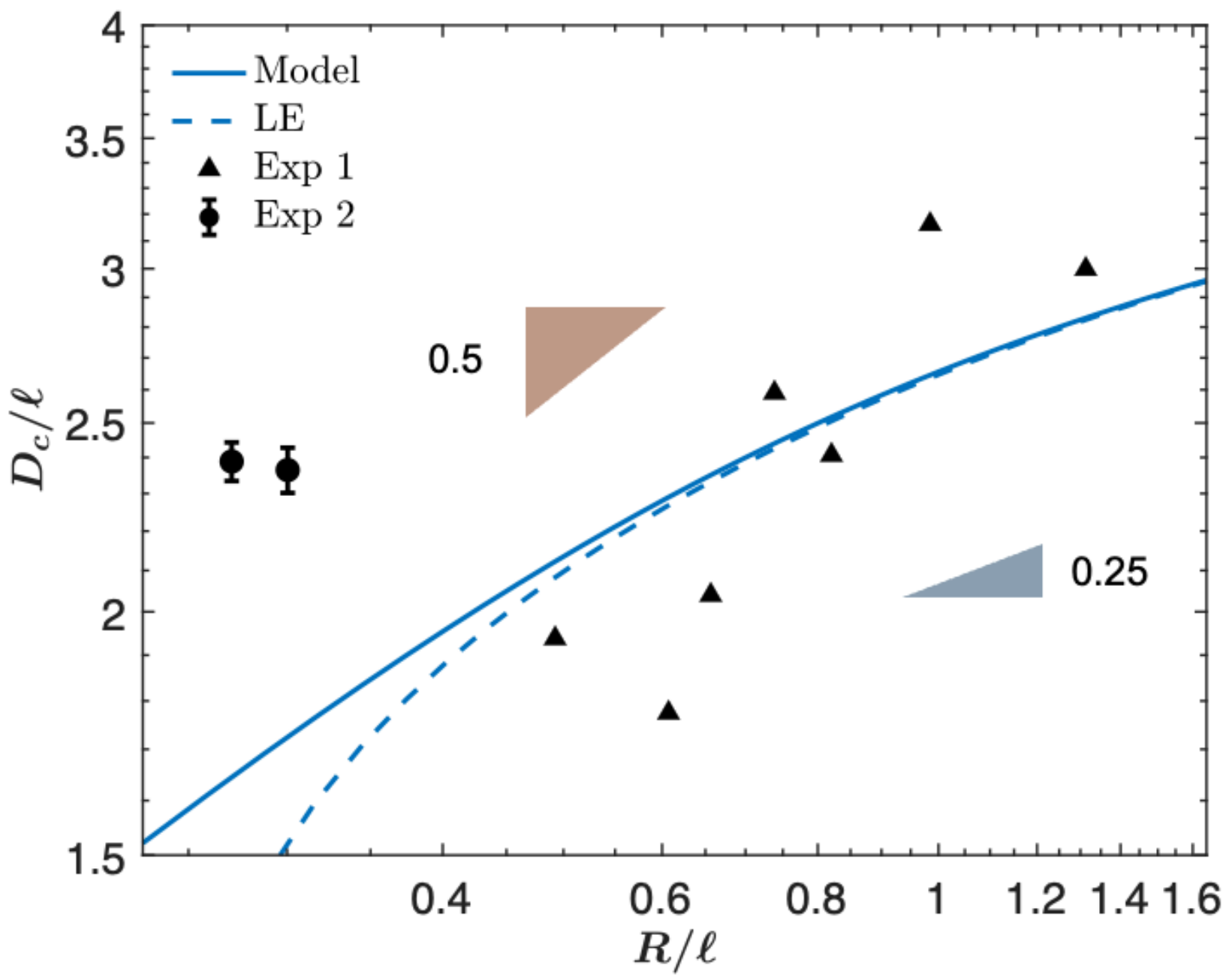

**Figure 6**: Dimensionless critical indentation depth at needle insertion, $D_c/\ell$, as a function of dimensionless needle radius $R/\ell$. Predictions from the nonlinear elastic model are shown by the solid line, while the linear-elastic approximation is shown by the dashed line. The predictions are compared with experimental data: Exp. 1 from [Lechenault2023] and Exp. 2 from the present study.

Fig. 7 reports the dimensionless critical insertion force $F_c/(\mu\ell^2)$ as a function of the dimensionless needle radius $R/\ell$ on logarithmic axes. Experimental measurements are compared with the nonlinear elastic model prediction (solid line), the corresponding linear-elastic approximation used to estimate $F_0$ (dashed line), and the prediction obtained by neglecting the energetic contribution of friction during insertion (dash-dot line), as in [Fregonese2021] for solid needles. At larger needle radii, the nonlinear and linear-elastic predictions converge and approach an approximately linear scaling, $F_c \sim \mu \ell R = \Gamma R$, as indicated by the green slope triangle. In this regime, the critical insertion force is controlled primarily by fracture toughness and needle radius. A similar linear dependence of the critical insertion force on needle radius was previously observed for solid-needle insertion [Fakhouri2015]. This scaling is also consistent with the behavior of the frictionless penetration force $F_0$, while the connection between $F_0$ and $F_c$ has been reported by [Fregonese2021, Fregonese2023]. Importantly, the frictionless prediction substantially underestimates $F_c$, whereas the full model including friction captures the experimental values well over most of the investigated range. This comparison shows that frictional dissipation during insertion provides a significant contribution to the critical force and is required for quantitative agreement with experiments.

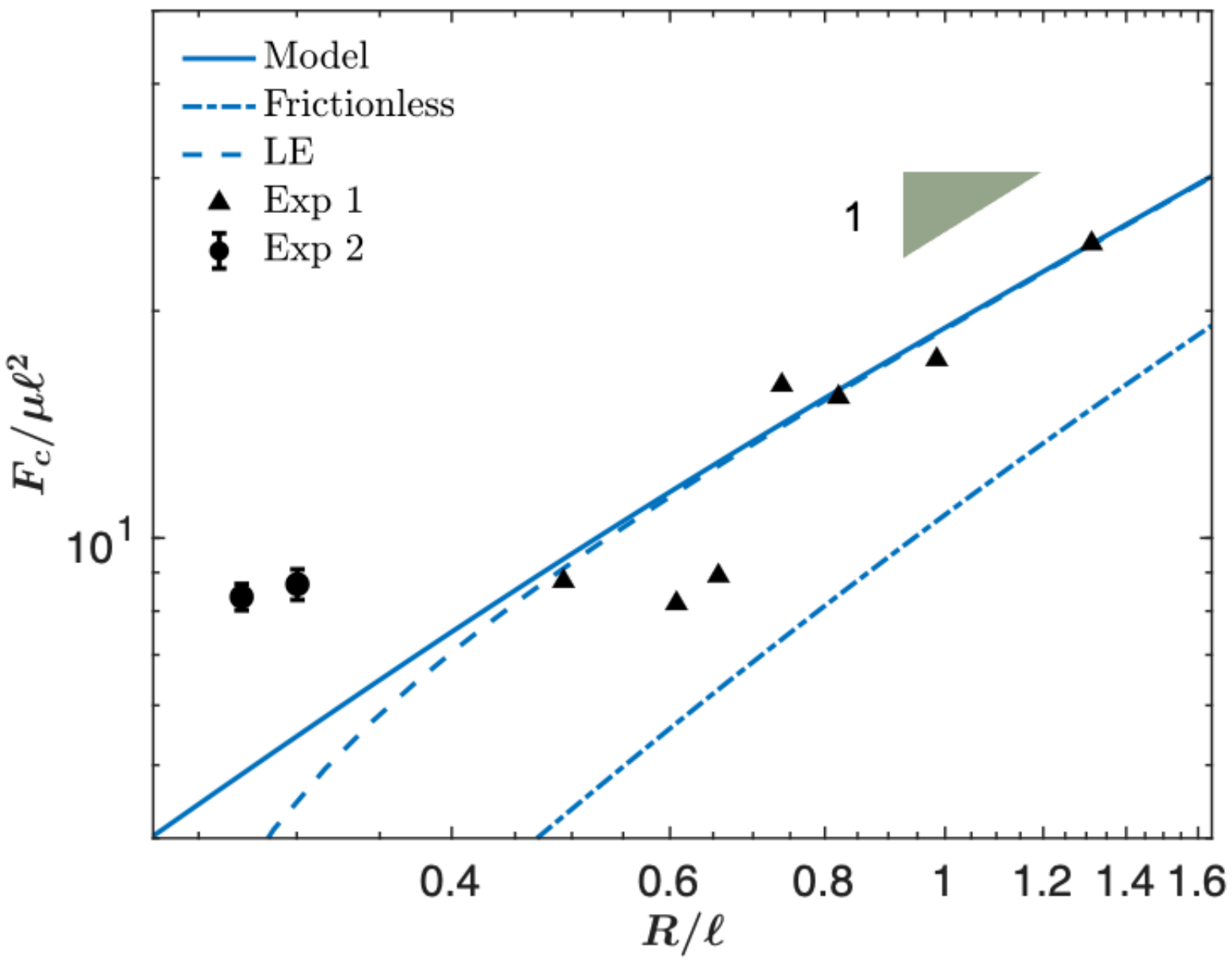

**Figure 7**: Dimensionless critical indentation force $F_c/(\mu \ell^2)$ at needle insertion versus dimensionless needle radius $R/\ell$. Predictions from the nonlinear elastic model (solid line) and the linear elastic approximation (dashed line) are compared with experimental data (see Fig. 3).

## Conclusions

This work develops an energy-based analytical framework for hollow-needle puncture that captures the coupled roles of fracture, elasticity, needle geometry, and friction. The model provides predictions for five key quantities governing puncture mechanics: the extracted core size, the frictionless penetration force, the frictional force slope during penetration, the critical insertion depth, and the critical insertion force. Comparison with experimental measurements shows that the model captures the main trends across a wide range of needle radii and provides a unified description of the hollow needle puncture process. In particular, comparison with the frictionless prediction shows that frictional dissipation during insertion is essential for quantitative prediction of the critical insertion force.

The scaling relations observed in Figs. 3-7 reveal how puncture mechanics is governed by the interplay between needle geometry and material properties. In particular, the results highlight the role of the elasto-fracture length $\ell = \Gamma/\mu$, which sets the balance between crack-propagation resistance and the elastic resistance to opening the crack. All predicted quantities increase with needle radius, but with different scaling exponents reflecting the distinct mechanisms governing each stage of the puncture process. The insertion forces exhibit a stronger dependence on fracture toughness than on elastic stiffness, indicating that puncture is primarily governed by the energetic cost of creating new fracture surfaces. In contrast, the frictional force slope shows a stronger dependence on stiffness, reflecting its relation to contact pressures along the needle shaft. The extracted core size increases with stiffness and decreases with toughness, as tougher materials require a higher energy release rate at the crack front, favoring smaller cores. The critical insertion depth decreases with increasing stiffness, since stiffer substrates accumulate elastic energy more rapidly during indentation.

The comparison with experiments also highlights secondary effects that are not fully captured by the idealized model. Remaining discrepancies may arise from rate- and specimen-size-dependent effects [Barney2024], finite needle-edge or process-zone effects, uncertainty in the independently measured interfacial parameters, and residual indentation after insertion, all of which can modify the partition between stored elastic energy, fracture energy, and frictional dissipation during penetration.

Beyond its mechanistic insights, the framework developed here provides a basis for predictive models of needle–tissue interaction. Accurate prediction of insertion forces is essential for needle-based medical procedures such as biopsy, drug delivery, and robotic needle insertion, where force feedback and trajectory control play a critical role. In particular, predictive models of puncture mechanics can inform trajectory planning, improve control strategies in robotic surgical systems, and support real-time simulations of needle insertion in soft tissues.

While the present study considers homogeneous, isotropic soft solids under quasi-static insertion conditions, the energetic framework established here provides a foundation for future extensions. Future work should also examine finite wall-thickness effects, since the present model applies only when the needle wall is sufficiently thin to avoid contact between the extracted core and the inner wall. Incorporating tissue heterogeneity, anisotropy, and rate-dependent fracture or frictional behavior will further improve predictive capability. Such developments may enable more reliable and adaptive models of needle insertion, ultimately contributing to improved accuracy, reduced tissue damage, and enhanced safety in minimally invasive biomedical procedures.

## Appendix A: Experimental Protocols

Figure A1 summarizes the experimental protocol used for the hollow-needle puncture tests. Samples of ED32 were placed on a wooden base plate to prevent needle damage after full penetration. The hollow needle was attached to the load cell through a mandrel and driven into the sample at a constant displacement rate. During the test, the axial force $F$ and penetration depth $D$ were recorded. After full penetration, the needle was retracted and the extracted core was recovered. The core was then imaged optically, and its diameter was measured along the stalk region, excluding the transient end regions. These measurements were used to determine the core radius $c$. Five repeated puncture trials were performed for each needle size, and the corresponding force–displacement curves are shown in Fig. A1e.

Figure A2 summarizes the experimental protocol used to characterize frictional resistance at the needle–sample interface. First, a cavity was created by hollow-needle puncture using a hole-forming needle of gauge $g_h$. The sample was then mounted on a custom stage that allowed a filler needle of gauge $g_f$ to travel through the full sample thickness $H$, while maintaining alignment through a guide hole. Filler needles of different diameters were inserted into the pre-existing cavity and then retracted while measuring the axial force $F$ as a function of displacement $D$. The plateau force $F_p$, measured when the filler needle was in contact with the sample over the full thickness $H$, was used to extract the frictional shear stress as $\tau_f = F_p/(2\pi\, R_f\, H)$, where $R_f$ is the outer radius of the filler needle. The corresponding contact pressure $p$ was computed from the cavity-opening model using Eq. (21). The resulting $\tau_f$–$p$ data were fitted with Eq. (19), allowing the pressure-independent shear strength $\tau_0$ and Coulomb friction coefficient $f$ reported in Table 1 to be identified. Table A1 reports the outer radii associated with the needle gauges used to create the hole, $g_h$, and to perform the friction tests, $g_f$.

**Table A1**: Outer radii associated with the needle gauges used in the friction tests.

| Gauge | $R$ $[mm]$ |
|---|---|
| $14G$ | 1.06 |
| $15G$ | 0.92 |
| $18G$ | 0.64 |
| $20G$ | 0.46 |

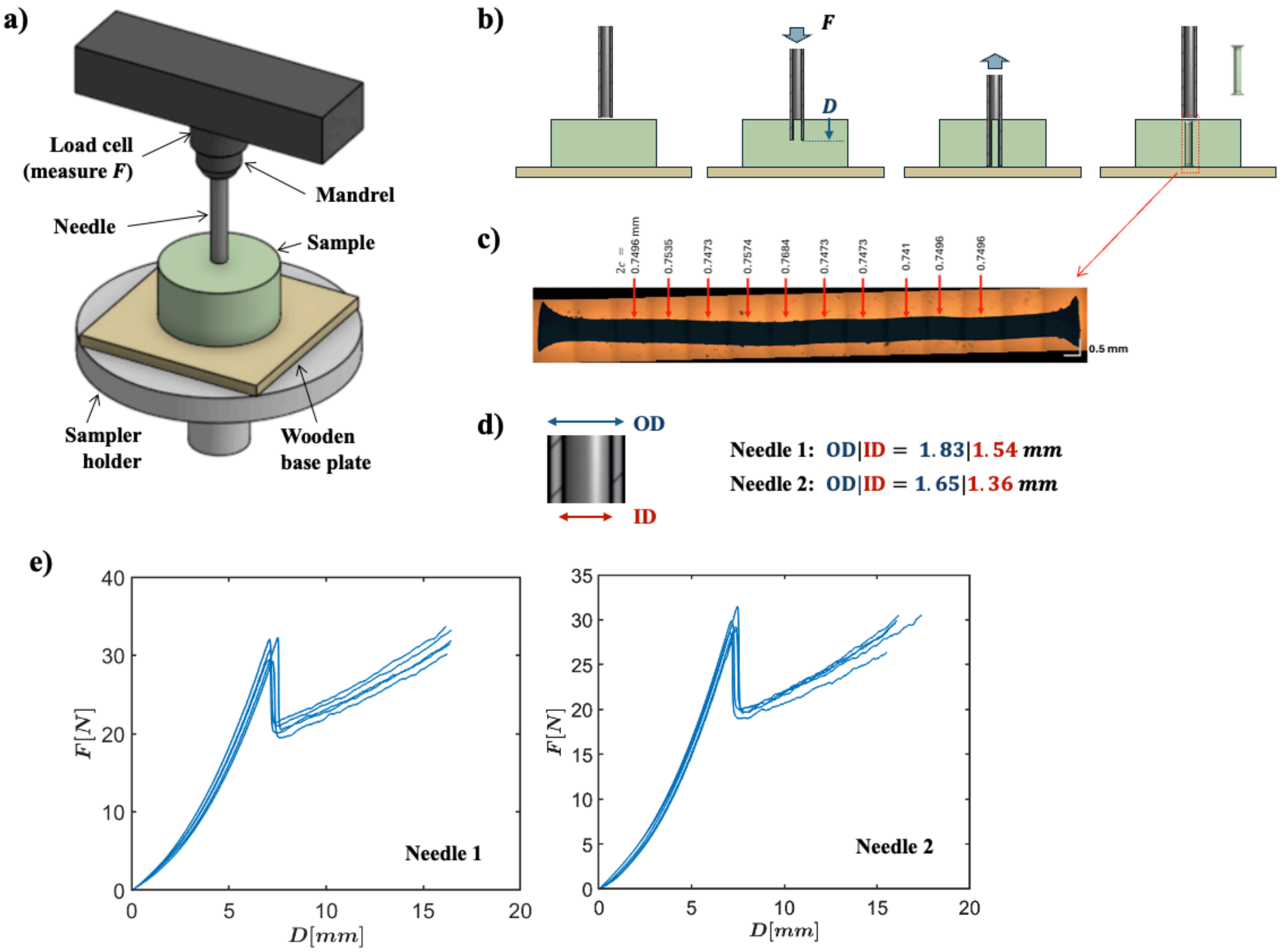

**Figure A1**: Experimental protocol for hollow needle puncture tests. **a)** Schematic of the puncture testing rig, including the load cell, mandrel, needle, sample, sampler holder, and wooden base plate. **b)** Schematic sequence of puncture, full penetration, needle retraction, and core extraction; the applied force $F$ and penetration depth $D$ are indicated. **c)** Representative extracted core obtained with Needle 1, showing image-based measurements of the core diameter $2c$, from which the core radius $c$ is obtained. **d)** Inner and outer diameters of the two hollow needles used in the experiments: Needle 1, OD/ID (=1.83/1.54) mm; Needle 2, OD/ID (=1.65/1.36) mm. **e)** Measured force–displacement curves for all five puncture trials using Needle 1, left, and Needle 2, right.

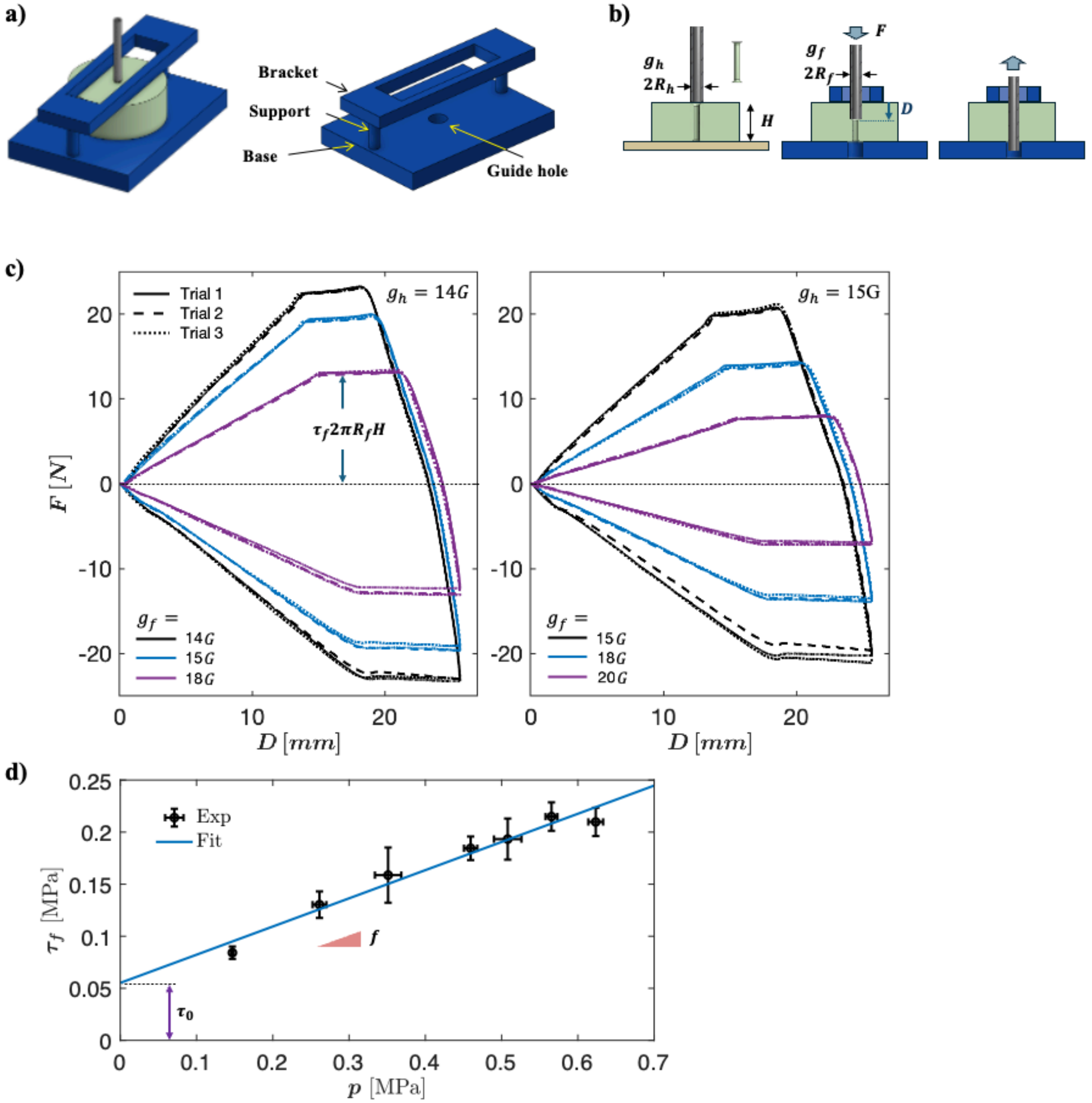

**Figure A2**: Experimental protocol and analysis for friction tests. **a)** Schematic of the custom stage used to secure the sample during needle reinsertion and retraction. The stage consists of a base, support, bracket, and guide hole to maintain needle alignment. **b)** Schematic sequence of the friction test. A cavity is first created by hollow-needle puncture using a hole-forming needle of gauge $g_h$ and corresponding outer radius $R_h$ (Table A1). A filler needle of gauge $g_f$ and corresponding outer radius $R_f$ (Table A1) is then inserted through the pre-existing hole and retracted while measuring the axial force $F$ as a function of displacement $D$. The sample thickness is denoted by $H$. **c)** Representative force–displacement curves for holes created using $g_h = 14G$ and $g_h = 15G$. Different colors indicate the gauge of the filler needle, $g_f = 14G, 15G, 18G, 20G$, while line styles indicate repeated trials. The force plateau associated with full-thickness contact is indicated as $\tau_f 2\pi R_f H$, where $\tau_f$ is the frictional shear stress. **d)** Frictional shear stress $\tau_f$ as a function of contact pressure $p$. Experimental data are fitted with Eq. (19), where the intercept gives the pressure-independent shear strength $\tau_0$ and the slope gives the Coulomb friction coefficient $f$.

## Appendix B: Theoretical findings

Fig. B1 reports the theoretical predictions of the model for the five main dimensionless quantities considered in this work: the normalized core size $c/\ell$, the frictionless penetration-force intercept $F_0/\mu\ell^2$, the post-insertion penetration-force slope $F_p'/\mu\ell$, and the critical indentation depth and force at insertion, $D_c/\ell$ and $F_c/\mu\ell^2$. These quantities are plotted as functions of the dimensionless needle radius $R/\ell$, where $\ell = \Gamma/\mu$ is the elasto-fracture length. To illustrate the sensitivity of the model predictions to material and interfacial parameters, we vary the Ogden strain-stiffening parameter $\alpha$, the Coulomb friction coefficient $f$, and the normalized pressure-independent friction $\tau_0/\mu$.

The first two quantities, $c/\ell$ and $F_0/\mu\ell^2$, depend only on the elastic-fracture part of the model and are therefore only affected by $\alpha$, and not by the frictional parameters. Increasing $\alpha$ increases both the predicted core size and the frictionless penetration force at fixed $R/\ell$. This reflects the larger elastic energy stored in the material surrounding the cavity for a higher strain-stiffening response. As $R/\ell$ increases, the predictions approach the linear-elastic asymptotic regime, where both quantities tend toward an approximately linear dependence on needle radius. At smaller $R/\ell$, nonlinear elastic effects become more pronounced and the dependence on $\alpha$ is stronger.

The last three quantities, $F_p'/\mu\ell$, $D_c/\ell$, and $F_c/\mu\ell^2$, depend on both the elastic-fracture response and the interfacial friction parameters. The penetration-force slope $F_p'/\mu\ell$ increases with both $f$ and $\tau_0/\mu$, as expected from the friction law used in the model. The pressure-independent contribution $\tau_0/\mu$ produces a direct increase in the interfacial shear strength, while the Coulomb contribution $f$ amplifies the effect of the contact pressure generated by cavity opening. Consequently, larger values of either frictional parameter increase the resistance to post-insertion penetration.

The critical insertion depth $D_c/\ell$ shows a more complex dependence because it is obtained from an energy balance between the elastic energy stored during indentation and the energy required to sustain penetration. Increasing friction generally increases the energetic cost of penetration and therefore increases the predicted critical depth. However, for high-friction cases, the dependence on $R/\ell$ may become non-monotonic, with a local maximum emerging from the competition between the frictionless penetration contribution, the frictional force slope, and the nonlinear indentation response. This behavior illustrates that the critical insertion depth is not controlled by a single scaling law across the full parameter range.

The critical insertion force $F_c/\mu\ell^2$ combines the effects observed in $D_c/\ell$ with the nonlinear indentation force-depth relation. It increases with $R/\ell$, $f$, and $\tau_0/\mu$, confirming that both needle size and interfacial friction raise the force required to trigger insertion. The dependence on $\alpha$ is mixed, because strain stiffening affects both the penetration energy and the indentation response. Overall, Fig. B1 shows that the five predicted puncture quantities are governed by different combinations of elasticity, fracture, geometry, and friction, explaining why different scaling trends are observed across the panels.

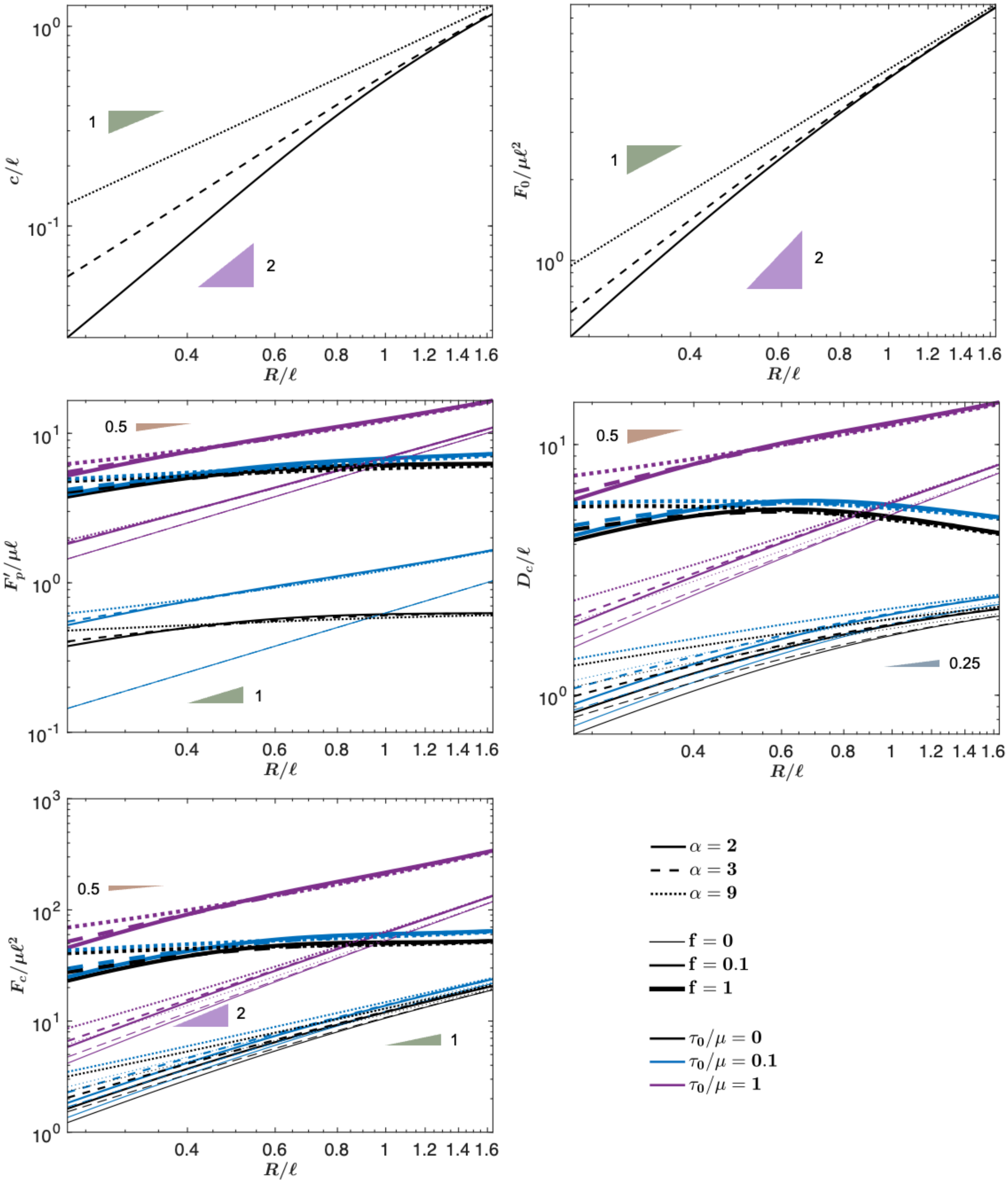

**Figure B1**: Theoretical predictions from the proposed model. The panels show the normalized core size $c/\ell$, frictionless penetration-force intercept $F_0/\mu\ell^2$, post-insertion penetration-force slope $F_p'/\mu\ell$, critical insertion depth $D_c/\ell$, and critical insertion force $F_c/\mu\ell^2$, plotted as functions of the dimensionless needle radius $R/\ell$. The strain-stiffening

parameter of the Ogden model is varied as $\alpha = 2,3,9$, while the interfacial parameters are varied as $f = 0,0.1,1$ and $\tau_0/\mu = 0,0.1,1$. Line style indicates $\alpha$, line thickness indicates $f$, and color indicates $\tau_0/\mu$. Slope triangles are included as visual guides to indicate approximate scaling regimes.

*Asymptotic Linear Elastic Expansion of $\varphi$*

Here we clarify the linear-elastic limit of the incompressible Ogden strain-energy density used in Eq. (7). For plane strain, $\lambda_z = 1$, and incompressibility gives $\lambda_r \lambda_\theta \lambda_z = 1$, so that $\lambda_\theta = \lambda_r^{-1}$. Setting $\lambda_r = 1 + e$, the Ogden strain-energy density becomes

$$\varphi = \mu \frac{2}{\alpha^2} [(1+e)^\alpha + (1+e)^{-\alpha} - 2] \tag{B1}$$

Now, expanding for $e \ll 1$, we have $(1+e)^\alpha \simeq 1 + \alpha e + \alpha(\alpha-1)e^2/2$, and $(1+e)^{-\alpha} \simeq 1 - \alpha e + \alpha(\alpha+1)e^2/2$, giving $(1+e)^\alpha + (1+e)^{-\alpha} - 2 \simeq \alpha^2 e^2$, and finally

$$\varphi \simeq 2\mu e^2 \tag{B2}$$

**Acknowledgments**

This work was supported by the Natural Sciences and Engineering Research Council of Canada (NSERC) (RGPIN-2025-07085).